\documentclass[preprint,aps,amsmath]{revtex4}
\usepackage{graphicx}
\def\fct#1{{\mathop{\rm #1}}}   
\def\sign{\fct{sign}}           

\begin{document} 

\title{ZM theory III: Classical oscillators and semi-classical Bohr-Sommerfeld quantization}

\author{Yaneer Bar-Yam}

\affiliation{New England Complex Systems Institute \\ 24 Mt. Auburn St., Cambridge, Massachusetts 02138}

\begin{abstract}
We consider the description of classical oscillatory motion in ZM theory, and explore the relationship of ZM theory to semi-classical Bohr-Sommerfeld quantization. The treatment illustrates some features of ZM theory, especially the inadequacies of classical and semi-classical treatments due to non-analyticity of the mapping of classical trajectories onto the ZM clock field. While the more complete ZM formalism is not developed here, the non-analyticities in the classical treatment resemble issues in the comparison of classical and quantum formalisms. We also show that semi-classical quantization is valid for a periodic manifold in ZM theory, though the quantum number $n=0$ is allowed, as it would be in quantum mechanics for a periodic manifold. Still, this suggests a connection to the first-order success of Bohr theory in describing the phenomenology of atomic quantum states. The approximate nature of the semi-classical treatment of three dimensional atomic orbits is, however, also apparent in relation to ZM theory. These observations are preliminary to a discussion of ZM theory in relation to quantum mechanics and quantum field theory in subsequent papers.
\end{abstract}


\maketitle

\section{Overview}   

In this paper we continue an investigation of the implications of a novel assumption about the relationship of fields, space and time. The central assumption is that time can be obtained from the values of a field variable. This treatment of the field variable $Z$ over the space manifold $M$ leads to ZM theory, which will be developed through a sequence of papers that explore the implications of such an assumption considering the compatibility of the approach with conventional physical laws, principles, mechanics, elementary forces and excitations. This paper is largely devoted to the description in ZM theory of systems that correspond to classical and semi-classical quantized oscillators. We continue the practice of being meticulous about providing details of derivations to clarify the assumptions. Since much of this repeats the conventional oscillator equations, this can be considered a ``worked out" exercise.

\section{Introduction}   

We review the basic framing of ZM theory from the first paper in the series, ZM1 \cite{ZM1}  with sufficient detail for this paper to be essentially self-contained. We consider a system, $Z$, with some set of distinctly labeled states that perform sequential transitions in a cyclic pattern, i.e. an abstract clock. Discreteness of the clock will not enter into the discussion in this paper. The clock states can therefore be extended to cyclical continuum, $U(1)$. The state change of the clock defines proper time, $\tau$, as defined by the clock. Since the clock is cyclical it is possible to represent the changing state using an oscillator language: 
\begin{equation}
\psi  = \exp ( - im \tau ), 
\end{equation}
where 
\begin{equation}
m = 2\pi /T
\end{equation}
is the cycle rate in radians. The clock phase is $c = m \tau$ modulo $2\pi$, though this expression is not analytic so that derivatives should be defined in terms of $\psi$.  However, where analytic continuation is valid, derivatives can be defined in terms of $c$.
The notation is chosen anticipating that $m$ will become the `rest mass' of the clock when it is reinterpreted as a particle. 
{\em A-priori} there is no difference between clockwise and counter-clockwise rotation, however, once one is identified it is distinct from the other.

To introduce the space manifold, $M$, we consider a parameter, $x$, associated with the environment such that properties of the environment may lead to variation of the clock state with $x$ (we assume $x$ is a real number parameter). The assumption of ZM theory is that observer time can be obtained by the rate of change of the clock, $c$, as a local gradient of the clock state in the two dimensional space given by $\hat \tau$ and $\hat x$ treated as a Euclidean space, whose direction can be considered as a `direction of time:'
\begin{equation}
(m, - k) =  m \hat \tau  - k\hat x.
\label{clock}
\end{equation}
We assume that time measures Euclidean distance along the time direction in the same units as $x$ and $\tau$, so the rate of change of the clock phase in the direction of time is 
\begin{equation}
\omega  = d_t c = \sqrt {m^2  + k^2 }, 
\label{omegatime}
\end{equation}
where $\omega$ is defined to be the magnitude of the rate of change of the clock phase.

We can specify locations in the two dimensional space by their coordinates along the axes 
\begin{equation}
(\tau_0,x_0)=\tau_0 \hat \tau + x_0 \hat x .  
\end{equation}
We can then write $t$ and $x$ in terms of $\tau_0$ and $x_0$. All expressions can be directly obtained from the geometry of the angle of time with space. For the case of uniform variation of clock time in space, and assuming time has no dependence on the observer defined space $x$, time is given by
\begin{equation}
t = (\omega/m) \tau_0.
\label{ttau00}
\end{equation}
The direction of time implies a shift in the position of the origin of $x$ toward negative values and a reference location translates to the right, with $x$ given by: 
\begin{equation}
\begin{array}{ll}
x&=x_0+(k/m)\tau_0 \\
&=x_0+(k/\omega) t.
\end{array}
\end{equation}
This identifies a connection between the velocity and $k$ in ZM theory. 

\section{Review of Hamilton's equations of motion}   

In paper ZM2 \cite{ZM2} we obtained Hamilton's equations of motion from the maximization of the rate of change of the clock with the direction of time. We will use these results here to consider a harmonic oscillator.  We review some of the key equations so that the application to the harmonic oscillator can be seen step by step in the formal treatment. Here we use conventional notation in advanced classical mechanics for the Hamiltonian $H$, momentum $p$, generalized space coordinate $q$, and velocity $\dot q$. These are otherwise associated with the variables $\omega$, $k$, $x$, and $v$ respectively.

The Hamiltonian as a function of $p$, is obtained as the maximum rate of change of the clock over the possible directions of time. The generalized function for arbitrary directions of time is 

\begin{equation}
H(p,\dot q) = p\dot q + m\sqrt {1 - \dot q^2 }.
\end{equation}
where the velocity $\dot q$ specifies the angle of time
\begin{equation}
\dot q = \sin(\zeta),
\end{equation}
where $\zeta$ is the angle of time from $\hat \tau$ in the $- \hat x$ direction.
Maximizing $H(p,\dot q)$ with respect to $\dot q$ gives 
\begin{equation}
p = m\dot q/\sqrt {1 - \dot q^2 }, 
\end{equation}
solving for $\dot q$ and inserting the result into $H(p,\dot q)$ gives 
\begin{equation}
H(p) = \sqrt {p^2  + m^2 }. 
\end{equation}
which is the Hamiltonian as a function only of momentum. Considering $p$ to be a variable function of $q$ we obtain the first of Hamilton's equations
\begin{equation}
 dH(p)/dp = \dot q.
\end{equation}
Dynamics is given by the time derivative of $p$, which was shown by differentiation to be
\begin{equation}
dp/dt = dH(p)/dq. 
\label{dpdtfirst}
\end{equation}
Assuming a conserved total energy leads to the definition of the potential energy
\begin{equation}
H(p,q) = H(p) + \phi (q), 
\label{spaceterminH}
\end{equation}
and
\begin{equation}
d\phi(q) /dq =  - dH(p) / dq .
\label{phidefine}
\end{equation}
Inserting this into Eq. (\ref{dpdtfirst}) gives
\begin{equation}
\begin{array}{ll}
dp / dt &=  - d\phi(q) / dq \\
&=  - \partial H(p,q) / \partial q, 
\label{Hamiltonssecond}
\end{array}
\end{equation}
which is the second of Hamilton's equations. 

Finally, defining the Lagrangian 
\begin{equation}
\begin{array}{ll}
L(q,\dot q) &= - m d _t \tau_0 - \phi (q) \\
&= p\dot q -  H(p,q)
\end{array}
\end{equation}
leads to the Lagrangian equations of motion as
\begin{equation}
(d/dt) \partial L(\dot q,q)/\partial \dot q = \partial L(\dot q,q)/\partial q.
\end{equation}

\section{Non-relativistic equations of motion}

For our study of the conventional harmonic oscillator we take the non-relativistic limit, the usual limit in which the oscillator is considered.  The non-relativistic limit of the equations corresponds to a small angle of deviation of the time direction to the direction of proper time. The generalized Hamiltonian as a function of the angle becomes: 
\begin{equation}
H(p,\dot q) = p\dot q + m(1- \dot q^2 /2).
\end{equation}
Maximizing $H(p,\dot q)$ with respect to $\dot q$ gives 
\begin{equation}
p = m\dot q, 
\label{pvsq}
\end{equation}
and 
\begin{equation}
H(p) = m+p^2 /2m, 
\end{equation}
which is the non-relativistic Hamiltonian. Other equations are unchanged from the relativistic case.

An interesting note, separate from the formal development, is that the linear dependence of $p$ on $\dot q$ in Eq. \ref{pvsq} in the non-relativistic case perhaps plays a role in obscuring the distinction between dependent and independent variables in the development of physical theories.

\section{Classical harmonic oscillator}

Our objective is to describe the behavior of a harmonic oscillator using the classical formalism treatment of ZM theory in which analytic continuation of the clock phase $c$ is assumed. In this case, it seems that describing system environment relationships in ZM theory, i.e. specific instances of clock variations in space, can start from an expression for the spatially dependent momentum $p(q)$. We will see later that this approach is not adequate, because analyticity is not possible for many such expressions, but it is helpful as a way of exploring basic concepts and revealing why additional analysis must be made to describe ZM theory and its phenomenology.

From the classical perspective, a wide range of expressions for $p(q)$, if not completely arbitrary, can be used. Choosing an expression for $p(q)$ will result in a set of trajectories and a corresponding potential energy from the Hamiltonian formalism. For example, anticipating the desired expression for a harmonic oscillator from traditional treatments, we consider the candidate expressions
\begin{equation}
p = \xi_1 m \omega_1 \sqrt{q_1^2 - q^2}.
\label{pdef}
\end{equation}
where $m$ is inserted for convenience of units, and $\xi_1$ takes one of the values $\pm 1$ in alternative candidate expressions. By Eq. (\ref{pvsq}) the velocity is 
\begin{equation}
\dot q =   \xi_1 \omega_1 \sqrt{q_1^2 - q^2}.
\end{equation}
and the Hamiltonian written as a function of $q$ is
\begin{equation}
H(p(q)) = m(1 + \omega_1^2 (q_1^2 - q^2)/2).
\end{equation}
We take the derivative to obtain
\begin{equation}
d H(p(q))/dq = - m \omega_1^2 q.
\end{equation}
This gives the potential energy using
\begin{equation}
d \phi(q)/dq = m \omega_1^2 q,
\end{equation}
to obtain
\begin{equation}
\phi(q) = \phi(0) +  m \omega_1^2 q^2/2,
\end{equation}
where $\phi(0)$ is an arbitrary constant. Rather than setting it to zero, we will keep it explicit.
The total energy is then:
\begin{equation}
H(p,q) =  m+p^2 /2m + \phi(0) +  m \omega_1^2 q^2/2.
 \end{equation}
Inserting the expression for $p(q)$ gives
\begin{equation}
H(p(q),q)  = m(1+\omega_1^2 q_1^2/2)+\phi(0),
\label{energy}
\end{equation}
which is independent of $q$, as it must be.

Dynamics is given by the derivative of $p$ with respect to time. From Hamilton's second equation, Eq. (\ref{Hamiltonssecond}),
\begin{equation}
dp/dt = - m \omega_1^2 q.
\label{poft}
\end{equation}
This equation can be written in terms of a single variable, in more than one way. We can invert Eq. (\ref{pdef}) to give $q$ in terms of $p$
\begin{equation}
q=\xi_2  \sqrt{q_1^2-p^2/(m \omega_1)^2}, 
\end{equation}
where $\xi_2$ can take either of the values $\pm1$, independently of $\xi_1$.  Then, we insert this into Eq. (\ref{poft}) to obtain a first order equation
\begin{equation}
dp/dt = - \xi_2 m \omega_1^2  \sqrt{q_1^2-p^2/(m \omega_1)^2} . 
\label{diffeq}
\end{equation}
Alternatively, inserting Eq. (\ref{pvsq}) into Eq. (\ref{poft}) gives a second order equation when we explicitly interpret $\dot q$ as the time derivative of $q$:
\begin{equation}
d^2 q/dt^2 = -  \omega_1^2 q
\end{equation}
The solution of the first order equation can be obtained by inspection as
\begin{equation}
p = \xi_3 m q_1\omega_1 \cos( \omega_1 (t-t_0))
\end{equation}
where $t_0$ is an arbitrary constant, and $\xi_3$ is another variable that takes one of the two values $\pm 1$. It is apparent that $\xi_3$ remains fixed when analytically continuing the trajectory across the four combinations of signs of $\xi_1$ and $\xi_2$. We can determine directly from Eq. (\ref{pdef}) that 
\begin{equation}
\xi_1 = \xi_3 \sign ( \cos( \omega_1 (t-t_0))),
\end{equation}
and by inserting into the left and right sides of Eq. (\ref{diffeq}) that:
\begin{equation}
\xi_2 = \xi_3 \sign(\sin( \omega_1 (t-t_0))).
\end{equation}
Since a negative $\xi_3$ can be absorbed by a shift of $t_0$, we can more simply write
\begin{equation}
p = m q_1\omega_1 \cos( \omega_1 (t-t_0))
\end{equation}
From this, and  Eq. (\ref{pvsq}), or from the solution of the second order equation, we can obtain the trajectory:
\begin{equation}
q = q_1 \sin(\omega_1  (t-t_0)).
\end{equation}
Significantly, we see that this solution implies continuous transitions between negative and positive values of $\dot q$ or $p$, i.e. analytic continuation of the classical solution requires both of the possible values of $\xi_1$ to coexist in a single trajectory. The interpretation of this result is important for the understanding of ZM theory, we will discuss this issue through the following notes. 

The following comments explore both specific and general aspects of both the features and apparent limitations of the classical formalism as an approximate treatment of ZM theory. In particular, they discuss issues of analyticity as well as the existence of multiple values of the momentum at a particular location. At this stage they are interpretations and ultimately not essential to the formalism. Still, they can guide subsequent development of the formalism. 

First, it may be important to emphasize that the classical formalism approximation of ZM theory, which assumes that analytic continuation allows derivatives to be described in terms of $c$ rather than $\psi$, may not hold for the oscillator.

Second, the classical treatment of the oscillator above follows closely the conventional treatment of Newtonian mechanics. However, this approximation does not account for several features of ZM theory. Among the issues that are manifest are (a) the relationship of the given values of $p(q)$ to the behavior of ZM theory outside of the range of $q$ specified $q \in [-q_1,q_1]$, and (b) the classical analytic continuation between different values of $p(q)$ specified at a single location by $\xi_1$. 

(a) is problematic because we might expect that ZM theory requires $p(q)$ to be specified for all values of $q$ by a real value, because it represents the variation of the clock with coordinate $q$. If we extend the given value continuously (but non-analytically) by setting $p(q)=0$ everywhere except  $q \in [-q_1,q_1]$, we would arrive at the conclusion that we have a stationary particle everywhere except  in the region $q \in [-q_1,q_1]$ where it moves to the right(left) when $\xi_1$ takes the value $1(-1)$ respectively. This is not what we consider to be a classical oscillator, since the classical motion is confined to the oscillator region. More generally, we see that thus far we do not have a concept of confined motion in the treatment of ZM theory.

(b) is problematic because ZM theory appears to require a unique value of $p(q)$ at each value of $q$. Thus, the classical analytic continuation seems to be in direct contradiction with the formalism. This is one example of a more general concern that ZM theory may only allow one value of the velocity at any location. It might be hard to argue that existing experimental observations can definitely rule this out, as coexistence of two objects is not possible, and identifying a case where two objects cross the same actual location $q_0$ at different times according to the theory might be difficult. Still, the association of one velocity with one location in space conflicts significantly with conventional physical theories and should be discussed carefully before ZM theory could be accepted.

Third, we note that these two issues have parallels in the quantum mechanical treatment of a harmonic oscillator as well as the more general treatment of quantum systems. Both issues therefore suggest that the classical description of the motion does not adequately describe ZM theory especially at the points of momentum reversal. At least two features of quantum mechanics are relevant:  

(a) The quantum mechanical treatment deviates significantly from the classical treatment in the existence of tails of wavefunctions into the negative kinetic energy ``forbidden" regions beyond the range of classical motion. Moreover, the quantum wavefunction is defined everywhere in space, with localization a property of observation probabilities.

(b) A quantum mechanical wavefunction allows superposition which is used to describe oscillatory motion through the coexistence of partial waves moving in opposite directions. This quantum superposition is not the same as classical trajectories. The classical motion can be inferred by a classical limit correspondence but it is not the same as the quantum superposition. 

Thus we see that the existence of non-analyticities in considering the classical treatment of ZM theory has precedent in quantum mechanics, and suggests we consider more carefully the treatment of ZM theory to see whether it will further correspond to quantum mechanics. We will pursue this in subsequent papers.

Fourth, we see that in ZM theory, momentum reversal is a major singularity. The existence of a singularity is a warning that the classical approximation does not work in such circumstances. Thus, the concept of collisions must be understood carefully. The formal development introducing the potential energy suggests that potentials that arises from interactions should be understood as mean field approximations since they are introduced to account for the behavior of the system rather than describe the interaction directly.

Fifth, more specifically, the analytic continuation between $\xi_1 = \pm 1$, suggests that we are allowing the existence of multiple values of the momentum and velocity at a particular location in space. In this context the existence of multiple values is associated with extrinsic parameters describing coupling of the system to other coordinates. In classical physics, momentum is conserved in the system as a whole. A varying particle momentum only arises when multiple particles are coupled.  In the classical oscillator treatment we do not include all of the coordinates that appear in the physical oscillator. Oscillators can occur when two particles are linked together, where both particles oscillate, so that strictly speaking there is no such thing as a one-dimensional oscillator. Similarly, in ZM theory, multiple coordinates must be coupled so that a variation in momentum along one coordinate is coupled to a variation of momentum along another coordinate. The potential arises due to an effective field treatment of interactions describing the coupling between multiple clock fields. We have not yet described such coupling. The current treatment of the harmonic oscillator implies that such coupling must include the possibility of a changes in clock state variation over space, where such a change can be extended to include momentum reversal.

Sixth, to provide an understanding of how momentum reversal might be achieved in ZM theory (without modification of the theory), we note that in ZM theory velocity and acceleration arise as a relationship between observer and the clock, corresponding to the more classical concept of an observer-system relationship. To discuss this we use the fourth comment in Section II of ZM1 which introduces the association of observers with regions of space, and observation as an extrapolation of the space-time coordinates. This perspective is similar to that adopted in general relativity. However, in ZM theory metric changes are associated with an angle/velocity rather than the smoother second order curvature/acceleration in general relativity. Thus, in ZM theory motion can be related to the process of space extrapolation by an observer. This extrapolation leads to the angle of space with $\tau$ and with time. Hence, the concept of a flat space is a result of mapping the curvature/angle of space into the momentum and energy, and describing it from that perspective. From the discussion of the singularity in ZM theory, the mapping cannot be analytic when there is momentum reversal. The extrapolation of space at this point must reverse direction and space is no longer a single valued function of location. This is not a fundamental problem for the extrapolation, as there is no reason to believe that an observer's extrapolation should be a single valued function of location. It may be tempting to think that for a harmonic oscillator space is extrapolated into a circle, however this is not the case as a circle would correspond to infinite momentum at the points of reversal.  It is nevertheless important to recognize that the concept of a having multiple values of $p(q)$ arising from a non-analytic multiple-sheet space-time is consistent in principle with the concept of an observer dependent space-time. A multiply sheeted space-time would enable the formalism to accommodate velocity reversal at a particular location.

\section{Additional oscillator quantities}

The above treatment considers the classical variables and trajectory. From the point of view of ZM theory, it is interesting also to consider other quantities. For example, the direction of time is given by: 
\begin{equation}
(m,-p)=(m, -\xi_1 m \omega_1 \sqrt{q_1^2 - q^2}).
\end{equation}
The rate of change of the clock in the direction of time  is given by (rest mass plus kinetic energy):
\begin{equation}
 \omega = m(1 + \omega_1^2 (q_1^2 - q^2)/2).
\end{equation}
The displacement of the coordinate trajectory along proper time is given by:
\begin{equation}
\begin{array}{ll}
\int dt / \omega &= \int d \tau_0 / m \\
&= \Delta \tau_0 / m
\end{array}
\end{equation}
Which can be written in terms of the position coordinate using the expression for velocity rewritten as
\begin{equation}
d q =   \xi_1 \omega_1 \sqrt{q_1^2 - q^2} dt,
\end{equation}
giving:
\begin{equation}
\frac{dt }{ \omega}= \frac{ \xi_1 d q}{m(1 + \omega_1^2 (q_1^2 - q^2)/2) \omega_1 \sqrt{q_1^2 - q^2}} 
\end{equation}
To  first order in the non-relativistic approximation
\begin{equation}
\frac{dt }{ \omega}= \frac{ \xi_1 (1 - \omega_1^2 (q_1^2 - q^2)/2) d q}{m \omega_1 \sqrt{q_1^2 - q^2}} .
\end{equation}
Integrating gives the classical trajectory through the dependence of proper time $\tau$ on coordinate $q$:
\begin{equation}
\tau_0(q) =  \omega_1^{-1} \sin^{-1}(q/q_1)-\omega_1/2\left(q\sqrt{q_1^2-q^2}+q_1^2 \sin^{-1}(q/q_1)\right) .
\end{equation}
If we keep only the first order term in the limit of small $\omega_1$ it reduces to $t(q)$, 
\begin{equation}
t(q) =  \omega_1^{-1} \sin^{-1}(q/q_1),
\end{equation}
as is to be expected in the extreme classical limit.

In Fig. 1 the direction of time is plotted as a vector field along with the trajectory $\tau_0(q)$ in space-proper time coordinates. Note that it is apparent from this plot that the classical description of motion is insufficient because a time vector in the proper time direction, found at the limits of classical motion, implies that there is no variation of the clock field in space. Such a clock field would not be confined (i.e. ``localized" in a quantum sense) to the region of the harmonic oscillator. 

\begin{figure}
\includegraphics[clip=true, viewport=.0in 1.0in 3.0in 10.0in,width=5cm]{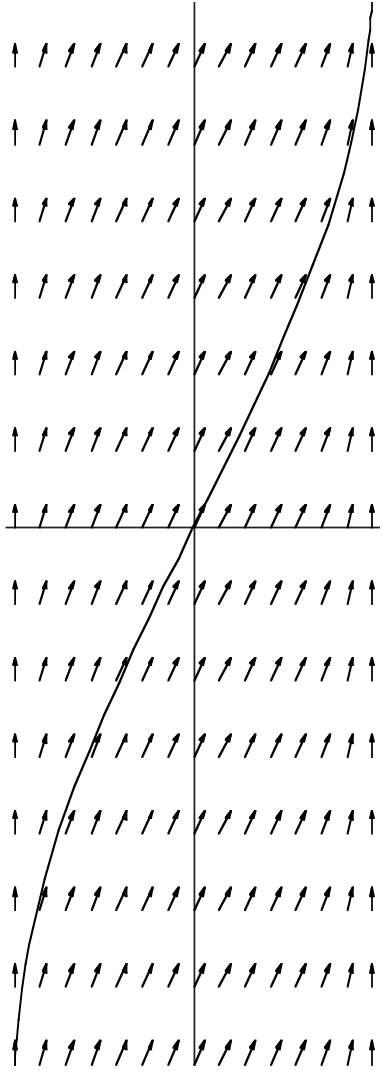}
\label{fig1}
\caption{Illustration of the clock state in space-proper time dimensions, and the trajectory $\tau_0(q)$ for a harmonic oscillator. We show only the case of $p>0$ at every location. The case of $p<0$ is obtained by inversion of space.  }
\end{figure}

\section{Bohr-Sommerfeld quantization}

The semiclassical Bohr-Sommerfeld quantization condition for the integral around orbits:
\begin{equation}
\int p dq = 2 \pi n
\end{equation}
where $\hbar$ is set to one, can provide a starting point for understanding quantization in ZM theory as it does in conventional modern physics, when applied in an approximate way to orbital motion. However, this condition does not describe all of the physics of orbital motion as described by quantum mechanics and quantum field theory.
Here, we show that in a periodic one-dimensional space the Bohr-Sommerfeld condition would be exact in ZM theory. It does not apply to periodic motion in space such as described by a harmonic oscillator due to the discontinuity described previously. 

In ZM theory the integral of $p$ over $q$ is given by:
\begin{equation}
\begin{array}{ll}
\int p dq &= \int d_{\hat q} c dq \\
&= c(q_f) - c(q_i)
\end{array}
\end{equation}
If we assume a periodic space so that the value of $c(q_f)$ is the same as $c(q_i)$ when including the periodicity of the clock phase, we have:
\begin{equation}
c(q_f) - c(q_i) = 2 \pi n
\end{equation}
with $n$ an integer, proving the Bohr-Sommerfeld condition for a periodic space. In contrast with the conventional Bohr-Sommerfeld condition, here we have no a-priori restriction that would prevent the case $n=0$. We note that for a quantum treatment of atomic orbits, the orbital component of the wavefunction can be constant (for an s-state) and the lowest energy of the first quantum state results from the radial component of the wave equation.

Conceptually, it is interesting to consider the application of this result to the oscillator \cite{sommerfeld} by considering the integral as a line integral over the classical trajectory with $dq$ positive (negative) for $\dot q$ positive (negative), i.e. for $p$ positive (negative) respectively. For each of the different branches
\begin{equation}
\begin{array}{ll}
\int p dq &= \int  m \omega_1 \sqrt{q_1^2 -  q^2} dq \\
&= \pi m \omega_1 q_1^2 /2.
\end{array}
\end{equation}
Combining the two branches and applying the quantization condition above:
\begin{equation}
 \pi m \omega_1 q_1^2 = 2 \pi n.
 \end{equation}
Inserting this equation into the expression for the energy Eq. (\ref{energy}) gives:
\begin{equation}
 H(p(q),q)  = m+n \omega_1+\phi(0).
\end{equation}
As is well known, this gives the correct size of the energy quantization, with $\hbar$ set to one,
but does not correctly describe the ``zero point" energy. Additional comments that are relevant to this discussion follow.

First, in quantum mechanics the result for the energy from the semi-classical approximation is corrected due to the need to describe the entire wavefunction. The same considerations will arise when we consider more fully ZM theory when describing the clock field in subsequent papers.

Second, more specifically, it is apparent that the semi-classical treatment does not include the full space time dependence of the clock field needed for ZM theory. Specifically, consider the degenerate case of an oscillator with momentum $p=0$ allowed by the quantization condition for $n=0$. The only description of such a system in this approximate description of ZM theory would have $p=0$ everywhere. This corresponds to a stationary particle everywhere in space, and is therefore not a reasonable description of the limiting case of an oscillator. We anticipate that a more complete description of the ZM clock field would still lead to a closure condition forcing quantization, but one that is different from the Bohr-Sommerfeld one. For example, we could obtain the correct energy quantization by starting from the condition 
\begin{equation}
\int p dq = 2\pi(n+1/2). 
\end{equation}
Making such a change must be justified by a more complete discussion of ZM theory. 

Third, while we have not yet extended ZM theory to three dimensions, it is reasonable to consider that in a three dimensional formulation the Bohr-Sommerfeld condition will apply at some level of approximation to particles in orbits. The approximation should be valid when the natural length scale in ZM theory, the size of the proper time dimension periodicity $1/m$, the Compton wavelength, is small compared to the size of the orbit. This is consistent with empirical observations, Bohr atomic theory, and quantum mechanics. It is also consistent with the existence of  corrections (e.g. the Lamb shift) found in quantum electrodynamics, considered as an expansion in the fine structure constant given by the ratio of the Compton wavelength to the first Bohr radius.

Fourth, the case of a particle in a box, given by 
\begin{equation}
p = \pm p_1 
\end{equation}
with $p_1$ a constant, and this value applying within a box specified by $q \in [0,q_1]$ appears to be closer to the case of a periodic space. It also gives the correct quantum energies for a closed line integral for cyclical movement of twice the length of the box. Still, strictly speaking, this does not correspond to the quantization condition of a periodic space, because $p=0$, i.e. $n=0$, is not allowed. A quantum treatment of a periodic space would also allow a $p=0$, $n=0$ solution. Thus it is clear that obtaining the minimum energy restriction for the particle in a box requires a more complete treatment that takes the box and its effect on the clock field explicitly into consideration.

Fifth, similar to the discussion of momentum reversing collisions for the case of the harmonic oscillator, the particle in a box must also have a coupling between momenta of the particle and the box. Collisions at the box walls transfer momentum between particle and box and thus quantum mechanical superposition conventionally discussed for the particle, must also account for the box state / wavefunction in the proper limit.  Thus, despite the superficial relevance of the simple quantization to the particle in a box, we expect that the collisions with the box are still non-analytic in ZM theory, as they must be to reverse momentum.

Sixth, we note that the natural closure condition on field variations of the clock field in ZM theory  is equivalent to the semi-classical quantization condition used for counting of independent states, which is the area in (p,q) phase space. 

Seventh, it is interesting to note that the concept of a multiple sheeted space that can separate intersecting trajectories was discussed by Einstein in 1917 in order to generalize Bohr-Sommerfeld quantization to multiple dimensions prior to the development of quantum mechanics.\cite{einstein} The subsequent development of quantum mechanics then avoided considerations of a multiple sheeted space. More recently Einstein's paper has been of historical interest because of its early recognition of the distinction between integrable and non-integrable motion.\cite{stone} In ZM theory the multiple sheeted space will continue to be relevant when we discuss its correspondence to quantum mechanics and quantum electrodynamics in subsequent papers.

I thank Marcus A. M. de Aguiar for comments on this and previous ZM manuscripts and for pointing out references \cite{einstein} and \cite{stone}.

\end{document}